# Advent of a gigawatt-class, tabletop, isolated-attosecond-pulse light source

BING XUE,[1,2] KATSUMI MIDORIKAWA,[1] AND EIJI J. TAKAHASHI[1,*]

[1]*RIKEN Center for Advanced Photonics, RIKEN, 2-1 Hirosawa, Wako, Saitama 351-0198, Japan*
[2]*bing.xue@riken.jp*
*Corresponding author: ejtak@riken.jp*

**We have produced a soft x-ray supercontinuum with a pulse energy of 0.24-µJ using high-order harmonics from a multi-terawatt, 10-Hz, stable, three-channel waveform synthesizer [Sci. Adv. eaay2802 (2020)]. We report here an attosecond streaking scheme, which is designed for measuring low-repetition-rate, high-energy, isolated attosecond pulses. The retrieved pulse duration is 226 as, which conclusively demonstrates the realization of 1.1-gigawatt isolated attosecond pulses at the generation point. The gigawatt-level peak power of this tabletop, isolated-attosecond-pulse light source is equivalent to that of an x-ray free-electron laser in a large facility.** © 2020 Optical Society of America

## 1. INTRODUCTION

Over the two decades since the first production of isolated attosecond pulses (IAPs) in 2001 [1], IAP sources have become highly desired to open up the broad research area of ultrafast science [2–4]. In particular, the high-power IAP sources that support attosecond-pump / attosecond-probe experiments are powerful tools for exploring ultrafast phenomena in the attosecond region [5–7] that cannot be accessed by a conventional fs-IR-pump / attosecond-XUV-probe scheme. One promising method for obtaining a high-power, gigawatt-level IAP is an X-ray free-electron laser (XFEL) [8–11] that utilizes the accelerator technology in a large facility. However, the stability of ultrashort pulses from XFEL sources is still limited—even in a very recent report [12]—owing to its amplification principle: self-amplified spontaneous emission [13]. Instabilities in the pulse duration, pulse energy, and spectrum shape also reduce the potential applications of XFEL sources to ultrafast dynamics research. Further, the accessibility of XFEL facilities is very limited, due to their huge costs and to space limitations. The development of laser-based, tabletop, high-power IAP sources is thus essential for eliminating these limitations of XFEL sources [12].

Generally, IAPs generated using laser systems have low output power because of the limited power of the laser system and the low conversion efficiency of high-order harmonic generation (HHG) [14,15]. Compared to XFELs, however, laser-based IAP sources have the advantages of temporal coherence, stability, and the ability to generate shorter pulse durations [16]. Thanks to their relatively low equipment cost, laser-based IAP sources are thus more suitable for laboratory-scale applications, if their low-power limitations can be overcome. With the recent development of a high-power, low-repetition-rate, near-infrared laser system [17], and an enhancement effect found in HHG driven by waveform synthesis [18], the generation of intense IAPs through HHG [19] is expected to be possible. Previously, we reported utilizing the HHG process to generate a high-power IAP by using a high-energy driving-laser source with a multi-terawatt (TW), three-channel, optical-waveform synthesizer at a 10-Hz repetition rate [20]. As an illustration of the results, we demonstrated the use of a custom-tailored optical waveform, obtained using parametric waveform synthesis, to generate an intense continuum spectrum covering the range 50–70 eV using HHG. By tailoring the synthesized waveform, we were able to obtain both an enhancement of, and a blue/red-shift in, the generated continuum spectrum. In addition, the bandwidth of the cutoff continuum with sub-microjoule energy supports transform-limited (TL) pulse durations of sub-200 as [20]. However, it is not easy to evaluate precisely the pulse duration of the low-repetition-rate, high-energy supercontinuum, for reasons that we explain later.

Streaking methods [3] are most frequently used to characterize the temporal shape or duration of IAPs, with the frequency-resolved optical gating for complete reconstruction of attosecond bursts (FROG–CRAB) method [21,22] used for retrieving the spectral intensity and phase information from the streaking results. However, the characterization of the IAP duration in an experiment with a low-repetition-rate laser source suffers the huge barrier of an extremely long data-acquisition time. One reason for this is the low photon flux in the generated IAP, and a digital converter usually has to be used for photoelectron counting to acquire the photoelectron spectrum. When the trigger frequency is low, the data-collection time increases in inverse proportion to that frequency. Furthermore, when the data-acquisition time is longer, the stability of the generated IAP becomes more critical, which sets another barrier to IAP applications. To our knowledge, there has been no previous report of a successful streaking experiment at less than a 1-kHz repetition rate. Another reason is that there is no realization of IAP generation by a low-repetition-rate, multi-cycle laser. Consequently, streaking schemes have not been investigated so far for evaluating low-repetition-rate, high-energy IAPs.

To characterize the IAP duration experimentally from the generated supercontinuum, we achieved a groundbreaking attosecond streaking measurement at a 10-Hz repetition rate. After retrieval, we found the shortest pulse duration of the IAPs generated by the synthesizer to be 226 as, which confirms that the peak power of the generated IAPs exceeded 1 gigawatt (GW). To our knowledge, this is the first demonstration of a GW-scale IAP output in a tabletop environment.

## 2. EXPERIMENTAL METHOD

For the high-energy, low-repetition-rate waveform synthesizer [20] that produces the three component pulses from the multi-cycle optical parametric-amplifier outputs—the pump (800 nm, 20.3 mJ, 30 fs), the signal (1350 nm, 4.3 mJ, 44 fs), and the idler (2050 nm, 1.6 mJ, 86 fs)—the first important issue in applying the attosecond streaking method is to remove the intense driving pulse after the HHG, because the energy of this pulse is high enough to damage optical components such as metal filters and multilayer mirrors easily. In this work, we therefore designed a polarization-controlled power scheme to reduce the remaining energy of the driving pulses and purify the IAP from the synthesizer pulse. The detailed scheme is shown in Fig.1. The gating pulse (800 nm, 30 fs, CEP-stabilized [23]) for the streaking is separated from the linearly polarized (p-polarized) pump pulse by a 45-degree, 2-inch mirror with a 20-mm-diameter center hole. The polarization of the gating pulse is later converted to s-polarization through a half-wave plate (HWP). After passing through the delay-control devices, the gating pulse is recombined with the synthesizer output by another center-holed mirror. For this two-beam-path configuration, a Mach–Zehnder interferometer is easily constructed by introducing another He–Ne laser beam passing through the same path. By monitoring the interference fringes, we obtain the feedback signal that we apply to the delay-control device (piezo stage) to suppress the delay jitter between the pump and the gating pulse to less than 56 as in RMS. The details are shown in the Supplemental Materials. Then, by varying the polarization spatially using the periscope, we finally adjust the output of the waveform synthesizer to s-polarization, while the gating pulse remains p-polarized before the HHG gas cell.

The pump pulse, signal pulse with idler, and gating pulse are focused separately using three lenses (focal lengths: 4.5 m, 3.5 m, and 4 m), while they pass freely through the entrance and exit holes of the HHG gas cell. Argon (Ar) gas is introduced hydrostatically into the 10-cm-long interaction cell. Two beam splitters (BSs) are introduced 2.5 meters away from the gas cell as the driving-pulse power limiter. The BSs are made of Si substrates with NbN (niobium nitride) coatings [24]. After two Brewster-angle reflections (74.8° at 800 nm) from the BSs, the s-polarized synthesizer output pulses are drastically reduced in power from tens of millijoules to several microjoules, while the HHG pulse and the p-polarized gating pulse are effectively reflected by the BSs. After the BSs, we used a spatial filter with two holes (2-mm and 1.5-mm in diameter) to transmit the IAP and shape the gating pulse, respectively. To optimize the focusing position of the gating pulse to overlap with the IAP at the end, a convex-concave pair of lenses with center holes is introduced after the spatial filter, with the IAP freely passing through the center holes of the lenses. We used a circular metal foil (Al or Zr) mounted on a mesh to remove the remaining energy of the synthesizer pulses completely from the HHG pulse, while the gating pulse can easily pass through the mesh (86% transmission) outside the metal foil. Another HWP with a center hole is also introduced between the convex-concave pair of lenses to change the gating pulse to s-polarization, which is the same as the IAP polarization. Both the IAP and the gating pulse are finally focused on and interact with a 0.13-mm-thick neon gas jet at the entrance to the electron time-of-flight (e-TOF) device. They are both focused by a multilayer Mo/Si concave mirror with a 300-mm focal length. This mirror is specially designed to have a high reflection rate (13%) around the 60-eV region in order to extract the continuum cutoff region from the harmonic spectrum and to have a 52% reflection rate for the gating pulse. The details can be checked in Fig.2(d). The estimated continuum harmonic energy is approximately 13 nJ at the interaction point in the e-TOF device.

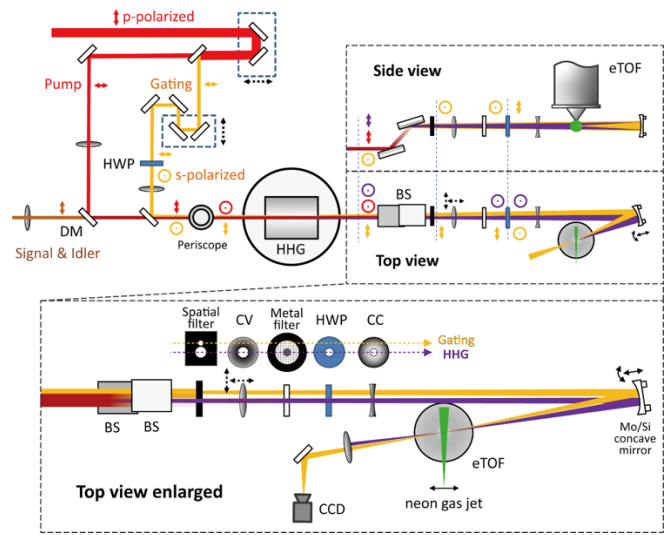

**Fig. 1.** Scheme of the system for attosecond streaking measurements. BS, beam splitter; HWP, half-wave plate; DM, dichroic mirror; HHG, gas cell for HHG; CV, convex lens with center

hole; CC, concave lens with center hole; eTOF, electron time-of-flight device.

Thanks to the high-energy continuum harmonics generated, for a sufficient flux of ionized electrons, the TOF signal can be observed directly with an oscilloscope instead of requiring a digital converter in counting mode. By getting rid of the digital converters, the data-acquisition time is greatly reduced, even at the 10-Hz repetition rate. A typical photoelectron spectrum measured with an Al filter is shown as the solid red curve in Fig. 2(d). We obtained this photoelectron spectrum by averaging 100 shots for 10 seconds.

## 3. RESULTS & DISCUSSION

We performed the attosecond streaking measurements by changing the gating-pulse delay using the piezo-controlled stage. The measured streaking trace is shown in Fig.2(a). The delay step of the gating pulse is set to 100 as. For each delay time of the gating pulse, 200 shots of photoelectron spectra measured with the e-TOF are collected and averaged by the oscilloscope. This streaking trace exhibits a clear modulation with a roughly 2.7-fs time period. This period corresponds to one cycle of the gating pulse, which has a center wavelength of 800 nm. This modulation lasts through the entire 20-fs scan range. It is related to the multicycle pulse duration (30 fs) of the gating pulse, and it can also be seen in the retrieved gating vector. We programmed homemade calculation software to retrieve the input pulses using the FROG–CRAB method with the principal-component generalized projections (PCGP) algorithm [25,26]. The retrieved trace shown in Fig.2(b) has a mean-square error of 3.5%. From the retrieved results for the IAP shown in Fig.3(c), we obtain an IAP duration of 226 as FWHM (full width at half maximum). By comparing the measured photoelectron spectrum and the Fourier transform (FT) spectrum obtained from the retrieved IAP profile, we find that the two spectra overlap quite well [see the inset in Fig. 2(c)]. Thus we conclude that this retrieved result is reliable. Accordingly, this confirms experimentally the generation of GW-scale IAPs (pulse energy of 0.24 μJ) using a fully stabilized, three-channel optical-waveform synthesizer [20]. As far as we know, this result is the first demonstration of an attosecond streaking experiment at a 10-Hz repetition rate, and it provides conclusive evidence for the realization of a GW-scale, soft x-ray IAP source on a tabletop.

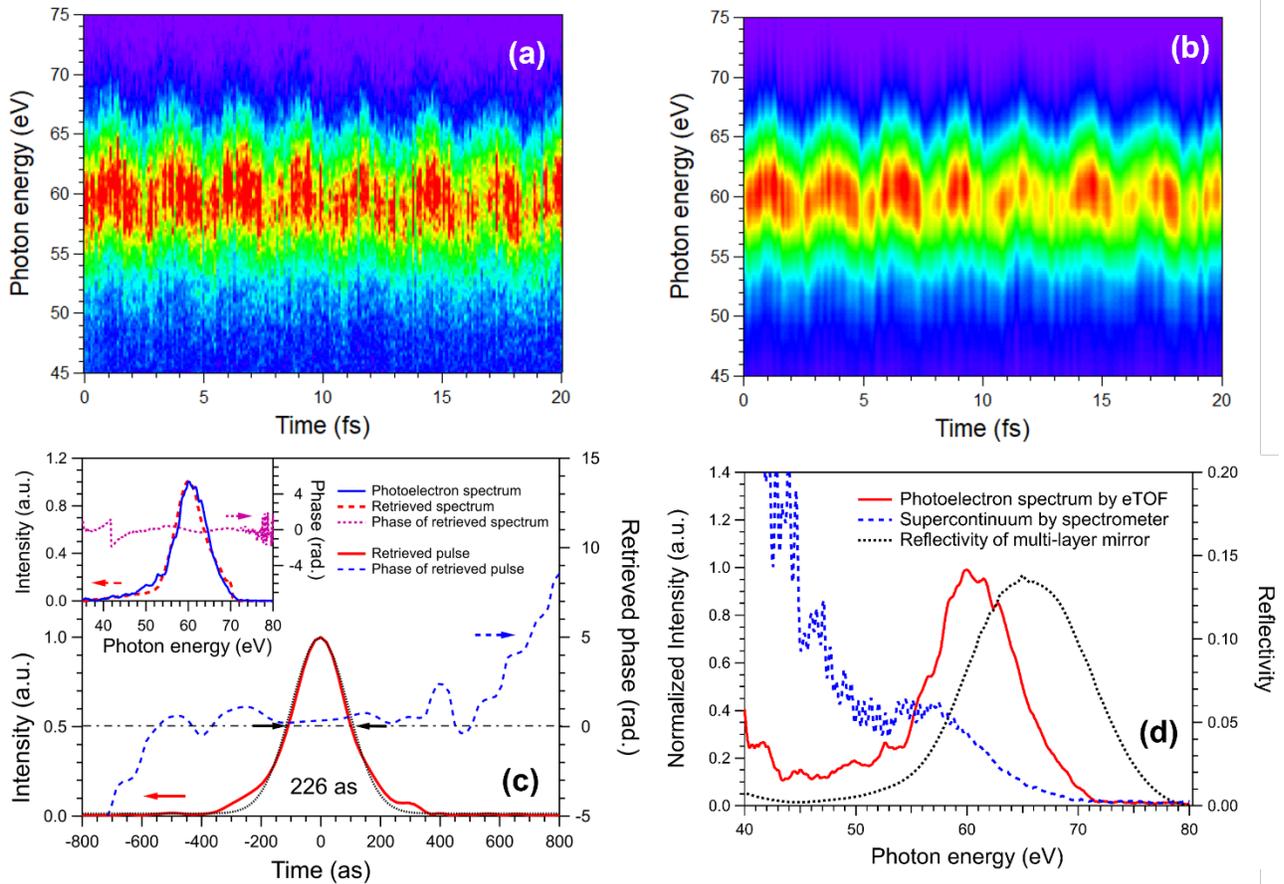

**Fig. 2**. (a) Experimental attosecond streaking trace and (b) the trace retrieved using FROG–CRAB with the PCGP algorithm; the retrieved FROG error is 3.5%. (c) The retrieved results for the pulse duration and phase of the IAP; the inset shows a comparison between the experimentally measured photoelectron spectrum and the FT spectrum from the retrieved pulse. (d) The e-TOF-measured photoelectron spectrum of the generated supercontinuum is shown as the solid red curve; the supercontinuum measured directly by the spectrometer is shown as the blue-dashed curve; and the reflectivity of the Mo/Si multilayer concave mirror is shown as the black-dashed curve.

As described previously, the cutoff of the generated spectrum can be red/blue shifted by changing the synthesized waveform used in driving the HHG process. Fig. 3(a) shows simulations of the shifted HHG spectrum when the delay time of the idler pulse in the synthesizer is changed by 0.8 fs [20]. The TL pulse duration is correspondingly changed from 236 as to 267 as, as shown in the inset in Fig.3 (a). We used the same Hanning window to filter the cutoff region of the spectrum during this calculation. Depending on the synthesized waveform conditions, the bandwidth of the measured photoelectron spectrum varies due to the combined effect of cutoff spectrum shifting and window filtering (in the experiment, this is done with the multilayer concave mirror and the metal filter).

In the experiment, when we changed the delay of the idler pulse in the synthesizer from an initial position A to a position B with a +1.7 fs delay, the photoelectron spectrum measured by the e-TOF device shifted from the red to the blue curve shown in Fig. 3(b). We applied four rounds of attosecond streaking sequentially with the delay positions located in the order A–B–A–B for the idler pulse of the synthesizer. From these attosecond streaking results, we find that the retrieved IAP durations obtained in the 1st and 3rd scans, with the idler-pulse delay at position A, are around 270 as [see Fig. 3(c)]. Correspondingly, for the 2nd and 4th scans, with the idler delay at position B, the retrieved IAP durations are consistently 240 as [see Fig. 3(d)]. These results provide strong evidence that stable and intense IAP generation has been achieved by the waveform synthesizer at a 10-Hz repetition rate. The duration of the output pulse can thus be tuned by tailoring the waveform synthesizer. In the current experiment, the shortest measured pulse duration (226 as) was limited by the bandwidth of the multilayer mirror, which is centered at 65 eV. It may be possible to overcome this bandwidth limitation by replacing this multilayer mirror with a grazing-incidence toroidal mirror.

The atto-chirp of the measured IAP is very small [27] because only the cutoff region of the spectrum is extracted by the multilayer mirror. This is also confirmed by the phase information retrieved for every streaking measurement.

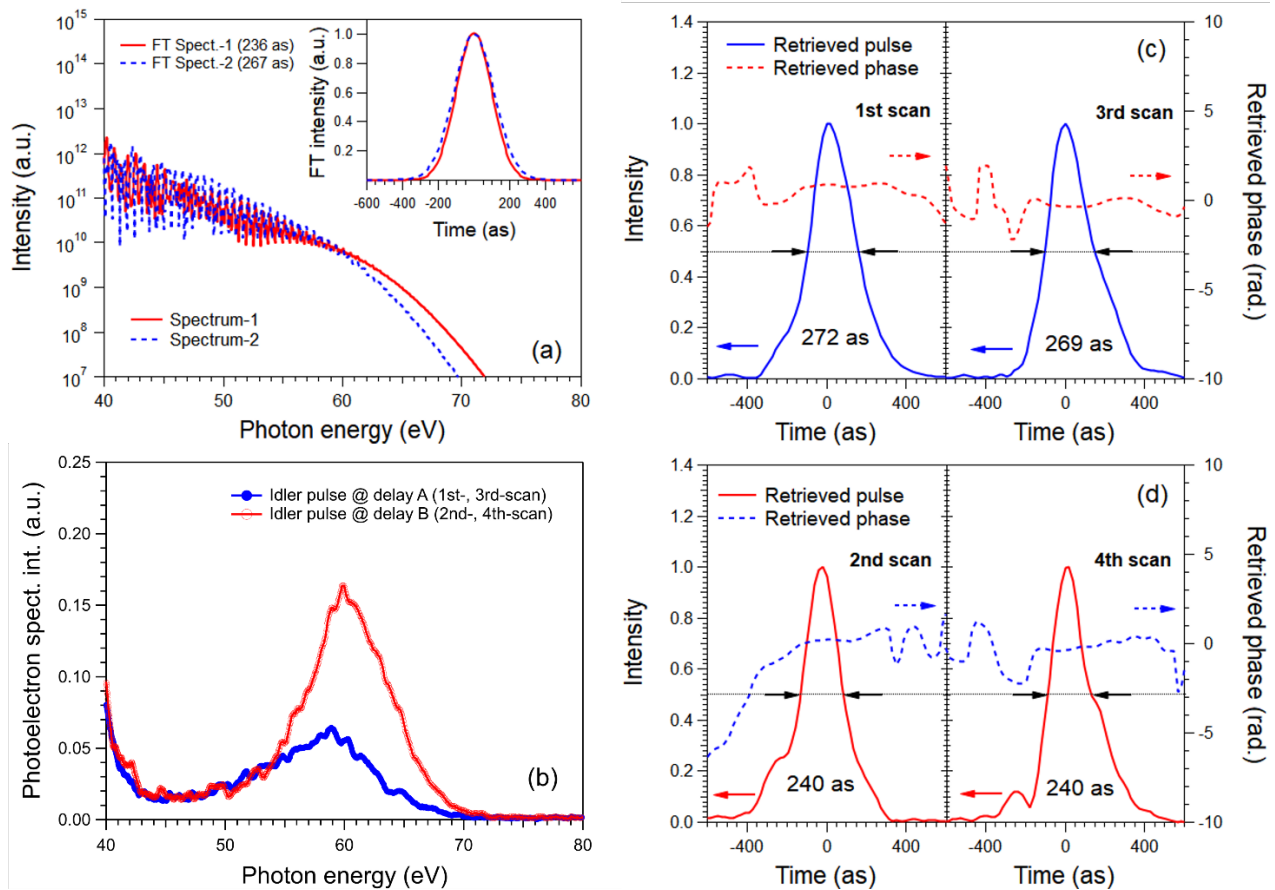

**Fig. 3.** (a) Simulated HHG spectrum–1 and –2, which correspond to a change of 0.8 fs in the delay of the idler pulse in the synthesizer. The inset in (a) shows the Fourier transforms of the pulses for spectrum–1 and –2. (b) Photoelectron spectrum of the IAP measured when the delay of idler pulse is changed by +1.7 fs (from delay-point A to B). (c) The retrieved IAP profiles with the idler pulse at delay point A (the 1st and 3rd scans) and (d) the corresponding profiles at point B (the 2nd and 4th scans).

## 4. CONCLUSION

We have demonstrated attosecond streaking measurements for characterizing a low-repetition-rate, intense IAP source. We confirmed experimentally the GW-scale IAP output (10 Hz, 0.24 µJ, 226 as) using the HHG from a waveform-tailored, high-power, multichannel synthesizer. The results prove that intense IAPs were generated, and the excellent stability of the synthesizer makes it possible to characterize the pulse duration by using the attosecond streaking method at a 10-Hz repetition rate. The streaking experiment also proves that tunability of the pulse durations of the IAPs generated by the synthesizer is achievable. Our streaking system can also be applied to measure the sub-µJ low-repetition-rate continuum HHG above 100 eV [28] by employing appropriate optics. The tabletop-based, intense IAP source developed in this work constitutes a breakthrough in the power and repetition-rate limitations of IAPs and paves the way for future applications [29–31] in attosecond science.

**Funding.** The Ministry of Education, Culture, Sports, Science and Technology of Japan (MEXT) through Grants-in-Aid under Grant Nos. 17H01067, 19H05628 and 21H01850. The MEXT Quantum Leap Flagship Program (Q-LEAP) No. JPMXS0118068681. FY 2019 President discretionary funds of RIKEN. The Special Postdoctoral Researchers' Program of RIKEN. The Matsuo Foundation 2018.

**Acknowledgments.** We thank Dr. Pengfei Lan and Dr. Hua Yuan for theoretical simulation support. We thank Dr. Yuxi Fu, Dr. Xu Lu, and Mr. Kotaro Nishimura for technical assistance. We thankfully acknowledge valuable discussions by Dr. Hiroki Mashiko and Dr. Katsuya Oguri.

**Disclosures**. The authors declare that they have no competing financial interests.

**Data availability.** Data underlying the results presented in this paper are not publicly available at this time but may be obtained from the corresponding authors upon reasonable request.